\documentclass[aps,prl,twocolumn,superscriptaddress]{revtex4}
\usepackage{amssymb}
\usepackage{amsfonts}
\usepackage{graphicx}
\usepackage{amsmath}
\usepackage{array}
\usepackage{lineno,hyperref}
\usepackage{footnote}
\begin{document}


\title{Enhancement of the superconductivity and quantum metallic state in the thin film of superconducting Kagome metal KV$_3$Sb$_5$}

\author{Teng Wang} \affiliation{State Key
Laboratory of Functional Materials for Informatics, Shanghai
Institute of Microsystem and Information Technology, Chinese Academy
of Sciences, Shanghai 200050, China}\affiliation{CAS Center for Excellence in Superconducting
Electronics(CENSE), Shanghai 200050, China}\affiliation{School of Physical Science and Technology, ShanghaiTech University, Shanghai 201210, China}

\author{Aobo Yu}
\affiliation{State Key Laboratory of Functional Materials for
Informatics, Shanghai Institute of Microsystem and Information
Technology, Chinese Academy of Sciences, Shanghai 200050,
China}\affiliation{CAS Center for Excellence in Superconducting
Electronics(CENSE), Shanghai 200050, China}\affiliation{University
of Chinese Academy of Sciences, Beijing 100049, China}

\author{Han Zhang}
\affiliation{State Key Laboratory of Functional Materials for
Informatics, Shanghai Institute of Microsystem and Information
Technology, Chinese Academy of Sciences, Shanghai 200050,
China}\affiliation{CAS Center for Excellence in Superconducting
Electronics(CENSE), Shanghai 200050, China}\affiliation{University
of Chinese Academy of Sciences, Beijing 100049, China}

\author{Yixin Liu}
\affiliation{State Key Laboratory of Functional Materials for
Informatics, Shanghai Institute of Microsystem and Information
Technology, Chinese Academy of Sciences, Shanghai 200050,
China}\affiliation{CAS Center for Excellence in Superconducting
Electronics(CENSE), Shanghai 200050, China}\affiliation{University
of Chinese Academy of Sciences, Beijing 100049, China}

\author{Wei Li} \affiliation{State Key Laboratory of Surface Physics and Department of Physics, Fudan University, Shanghai 200433, China}

\author{Wei Peng} \affiliation{State Key Laboratory of Functional Materials for
Informatics, Shanghai Institute of Microsystem and Information
Technology, Chinese Academy of Sciences, Shanghai 200050,
China}\affiliation{CAS Center for Excellence in Superconducting
Electronics(CENSE), Shanghai 200050, China}\affiliation{University
of Chinese Academy of Sciences, Beijing 100049, China}

\author{Zengfeng Di} \affiliation{State Key Laboratory of Functional Materials for
Informatics, Shanghai Institute of Microsystem and Information
Technology, Chinese Academy of Sciences, Shanghai 200050,
China}\affiliation{University
of Chinese Academy of Sciences, Beijing 100049, China}

\author{Da Jiang}\email{jiangda@mail.sim.ac.cn}
\affiliation{State Key Laboratory of
Functional Materials for Informatics, Shanghai Institute of
Microsystem and Information Technology, Chinese Academy of Sciences,
Shanghai 200050, China}\affiliation{CAS Center for Excellence in Superconducting
Electronics(CENSE), Shanghai 200050, China}\affiliation{University of Chinese Academy of Sciences, Beijing 100049, China}

\author{Gang Mu}
\email{mugang@mail.sim.ac.cn} \affiliation{State Key Laboratory of
Functional Materials for Informatics, Shanghai Institute of
Microsystem and Information Technology, Chinese Academy of Sciences,
Shanghai 200050, China}\affiliation{CAS Center for Excellence in Superconducting
Electronics(CENSE), Shanghai 200050, China}\affiliation{University of Chinese Academy of Sciences, Beijing 100049, China}

\vspace{10pt}

\begin{abstract}
Recently V-based Kagome metal attracted intense attention due to the emergence of superconductivity in the low temperature.
Here we report the fabrication and physical investigations of the high quality single-crystalline thin films of the Kagome metal KV$_3$Sb$_5$.
For the sample with the thickness of about 15 nm, the temperature dependent resistance reveals a Berezinskii-Kosterlitz-Thouless (BKT) type behavior, indicating the presence of two-dimensional superconductivity.
Compared with the bulk sample, the onset transition temperature $T^{onset}_{c}$ and the out-of-plane upper critical field $H_{c2}$ are enhanced by 15\% and more than 10 times respectively.
Moreover, the zero-resistance state is destroyed by a magnetic field as low as 50 Oe.
Meanwhile, the temperature-independent resistance is observed in a wide field region, which is the hallmark of quantum metallic state.
Our results provide evidences for the existence of unconventional superconductivity in this material.
\end{abstract}

 \maketitle
%
%
%
%

Materials with Kagome lattice manifest abundant exotic quantum phenomena including geometric spin frustration, non-trivial topological states, charge and spin density wave orders, etc~\cite{Norman2016,MaPRL2020,Kiesel2013}.
The discovery of superconductivity in the Kagome metal AV$_3$Sb$_5$ (A = K, Rb, Cs) added a new physical dimension to this novel system~\cite{Ortiz2019,Ortiz2020,Ortiz2021,Yin2021}.
Theoretically the unconventional pairing has been proposed for this material due to the
proximity to the multiple van Hove singularities of the superconducting (SC) state~\cite{wu2021arXiv}. In the experimental side, the double-dome-shaped evolution of the SC transition temperature $T_c$
with pressure has been confirmed by several groups.~\cite{Chen2021arXiv,ChenXu2021arXiv,Zhu2021arXiv}
Concerning the gap structure, the conclusions from different techniques are still inconsistent~\cite{Zhao2021arXiv,Duan2021arXiv}. The observation of zero-bias conductance peak
inside the vortex core indicates the possibility for the Majorana bound states~\cite{Liang2021arXiv}.
These experimental results initially show the signs of unconventional superconductivity. Currently more solid experimental results are necessary in order to confirm the unconventional behaviors of this material.
It is an important perspective to investigate the physical performances in low dimension.
Actually, in recent years, two-dimensional (2D) superconductivity has drawn great interest due to the emergence of new quantum phenomena,
including Ising superconductivity~\cite{Ising,Ising-2,Ising-3,Saito2016},
quantum metallic state~\cite{Bose-metal1,Bose-metal2,Bose-metal3,Bose-metal4,Qin2006}, Berezinskii-Kosterlitz-Thouless (BKT)
transition~\cite{BKT-1,BKT-2,Reyren1196}, and even the significant enhancement of $T_c$~\cite{Gozar,FeSe,TaS2-1,TaS2-2}.
Thus, revealing the performance of SC properties in the low dimension is vary
crucial in understanding the intrinsic properties of this material.

In this Letter, we report the mechanical exfoliation and superconducting properties
of the thin film of KV$_3$Sb$_5$. Both the onset transition temperature $T^{onset}_{c}$ and the out-of-plane upper critical field $H_{c2}$ are enhanced significantly in the thin samples.
Two-dimensional superconductivity is revealed by both the BKT-type resistance transition and the logarithmic evolution of the flux-flow activation energy.
Strikingly, a transition from the superconducting to quantum metallic state is induced by a very small perpendicular magnetic field ($\sim$50 Oe),
while the onset transition point is only moved to about 0.5 K by the magnetic field as high as 6000 Oe.
The present work reveals the unconventional behaviors in Kagome superconductors and also provides an important platform to study the dimensionality effect of this system.

The KV$_3$Sb$_5$ single crystals were grown by the
self-flux method~\cite{Ortiz2019}. Utilizing the characteristics of layered structure, the mechanical exfoliation method was adopted.
The KV$_3$Sb$_5$ thin flake is exfoliated from its single crystal by scotch tape and transferred onto SiO$_2$/Si substrate (SiO$_2$-300 nm; Si-500 $\mu$m)~\cite{Jiang2014}.
Typically the in-plane dimension of the fabricated samples can
be as large as 200 $\mu$m. All the studies in this work are carried out on samples with a thickness of 15 nm.
The electrical transport data were collected on the dilution refrigerator based on the physical
property measurement system (Quantum Design, PPMS) by a
standard four-probe method, and the silver paste was used to prefabricate the electrodes. The external magnetic field was applied perpendicular to sample surface and the electric current. The applied
electric current is 10 $\mu$A.

\begin{figure}\centering
\includegraphics[width=9.5cm]{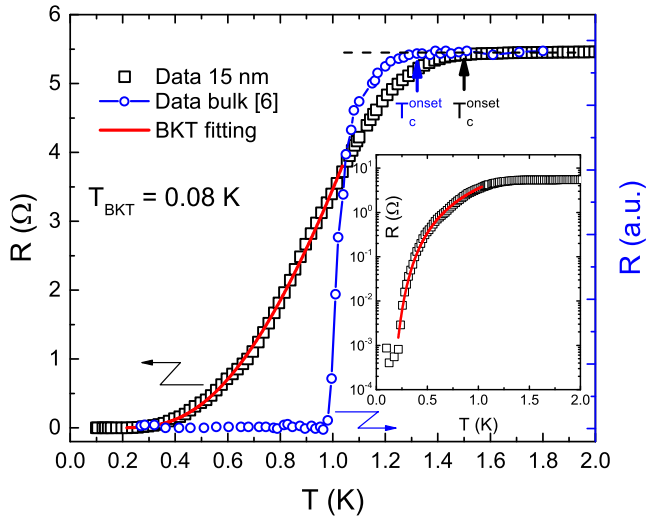}
\caption {(color online) Temperature dependence of resistance of KV$_3$Sb$_5$ thin film at zero magnetic field. The data of the bulk sample~\cite{Ortiz2021} is also shown for comparison (right axis).
The red line is the fitting curve representing the BKT transition (see text). Shown in the lower right inset is the same data using a semilog scale. } \label{fig1}
\end{figure}

Temperature dependence of resistance of the thin KV$_3$Sb$_5$ sample is shown in Fig. 1. The thickness of this sample is estimated to be 15 nm from the comparison of the resistance of the thin and bulk samples.
A SC transition is observed in the low temperature.
The onset transition temperature is enhanced by about 0.2 K as compared with that of the bulk sample~\cite{Ortiz2021}. Considering the rather low transition temperature in this material, actually this enhancement
reaches the extent as high as 15\%.
In sharp contrast, the zero-resistance temperature ($\sim$0.22 K) is much lower as compared with the bulk sample, revealing
a tail-like feature in a wide temperature range. Similar behaviors have been reported in other 2D SC materials, which were
explained in terms of the Berezinskii-KosterlitzThouless (BKT) transition~\cite{Reyren1196,Bose-metal4,Mo2C2015}. Based on this picture, the zero-resistance state is driven by the binding of vortex-antivortex pairs.
As shown by the red cure in Fig. 1, the experimental data can be well described by the equation~\cite{Halperin1979,Mo2C2015},
\begin{equation}\begin{split}
R=R_0 \exp[-b(\frac{T}{T_{BKT}}-1)]^{-1/2},\label{eq:1}
\end{split}\end{equation}
which demonstrates the occurrence of BKT transition. The BKT transition temperature is determined to be
$T_{BKT}$ = 0.08 K. The BKT-type behavior suggests preliminarily that 2D superconductivity is achieved at zero magnetic field in our sample. Considering the fact that the upper critical field is only several hundred osters
in the bulk sample~\cite{Ortiz2021},
the coherence length should be relatively large. Thus, it's rather reasonable that a film with the thickness of 15 nm can meet the requirement of two-dimensional limit.

\begin{figure*}\centering
\includegraphics[width=11.5cm]{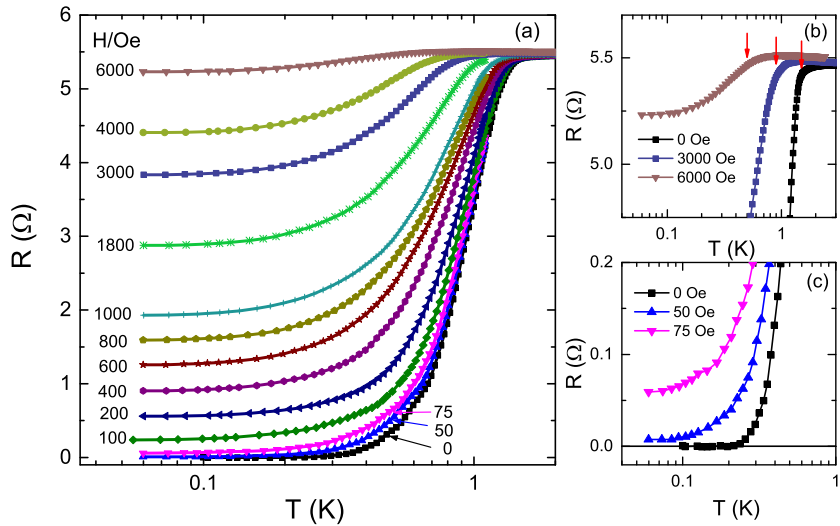}
\caption {(color online) (a) Electrical resistance as a function of temperature under the magnetic field up to 6000 Oe with $H$//c. (b, c) Enlarged view of the resistance data
in the vicinity of onset transition points and zero-resistance region respectively. The red arrows in (b) indicate the onset transition points.} \label{fig2}
\end{figure*}

We next focus on the electrical transport behaviors under out-of-plane magnetic field. As shown in Fig. 2(a), the $R-T$ curves show the systematically evolution with the increase of field.
These data provide three important messages that need to be emphasized in particular. Firstly, as shown in Fig. 2(b), the magnetic field as high as 6000 Oe suppresses the onset transition temperature  $T^{onset}_{c}$
(determined using the criterion 99\% $R_n$) from 1.5 K to 0.5 K, which indicates a rather high upper critical field above 6000 Oe. This value is more than 10 times higher than that of the bulk samples~\cite{Ortiz2021}.
Secondly, the zero-resistance state is destroyed by the magnetic field as low as 50 Oe, suggesting a very narrow flux solid state (if it exists). Thirdly, the temperature-independent resistance is observed in a
very wide field range from 50 to 6000 Oe, which is a hallmark of the quantum metallic state~\cite{Bose-metal1,Bose-metal4}. These striking features unambiguously demonstrate the dimensional effect of this system.

In order to further understand the unconventional temperature-independent behavior of the $R-T$ curves,
we plot the Arrhenius plots (ln$R$ vs $1/T$) under several typical fields in Fig. 3(a).
Ln$R$ shows a transformation from the linear evolution to the $1/T$-independent tendency with the increase of $1/T$, which are represented
by the dashed and dotted lines respectively. The intersections of the dashed and dotted lines are used to determine the transition temperature between the two behaviors.
The linear relation is the consequence of the thermal activated flux flow (TAFF) behavior,
where the resistance $R$ obeys the relationship~\cite{Vinokur1990,ZhangH2021}
\begin{equation}\begin{split}
R=R_0\exp(-\frac{U}{k_BT}),\label{eq:2}
\end{split}\end{equation}
where $k_B$ is the Boltzmann constant and $U$ is the thermally activated energy of the flux flow.
Thus $U(H)$ can be obtained from
the slope of this linear part in the Arrhenius plot. The acquired $U(H)$ at various magnetic fields from 50 to 6000 Oe are summarized in Fig. 3(b).
It is found that $U$ is proportional to -ln$H$ (see the dashed line in Fig. 3(b)). Theoretically it was pointed out that~\cite{FEIGELMAN1990177,White1993}, the activation energy is determined by the free energy
barriers to create a single free dislocation when the vortex
translational correlation length is small enough.
In this case, the logarithmic relationship can be expected. Thus our observations here strongly imply
the 2D liquid state for the vortices.

\begin{figure}\centering
\includegraphics[width=8.2cm]{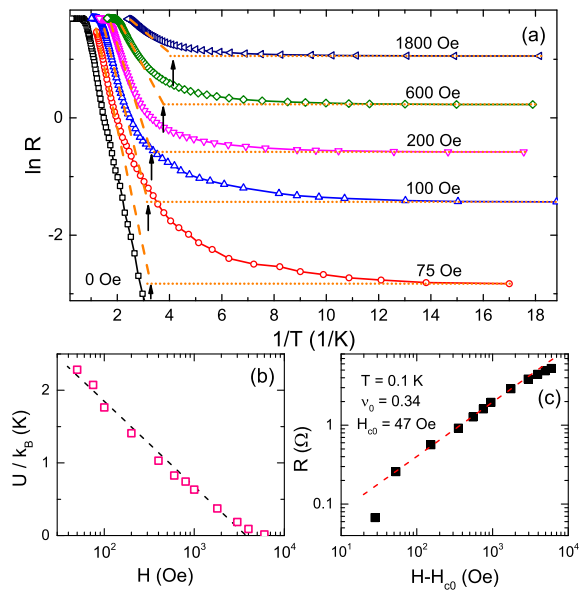}
\caption {(color online) (a) Arrhenius plots of longitudinal resistivity ln$R$ vs
$1/T$ at different fields. The data are shifted by 0.3 K$^{-1}$ for clarity. The dashed lines are the fits with TAFF theory to the
experimental data. The dotted horizontal lines are guides for eyes. The arrows indicates the transition points between the two regions with different behaviors. (b) The values of $U$ at various magnetic field.
(c) Magnetoresistance data at 0.1 K.}
\label{fig3}
\end{figure}

With the increase of $1/T$, the resistance becomes constant, which is independent of temperature. Similar phenomena have been reported in other 2D SC systems and are considered as a hallmark
of the quantum metallic state~\cite{Bose-metal1,Bose-metal2,Bose-metal3,Bose-metal4,Qin2006}. Such a metallic state in the vicinity of SC region could not be understood in the classic framework. Different theories have been proposed to interpret
this dissipation in the low temperatures. Typically the different models can be distinguished by the magnetic field dependent behaviors of the saturated resistance at low temperatures.
By considering the temperature-independent quantum tunneling of vortices, Shimshoni et al. predicted that the field induced increase of resistivity in the quantum metallic state should follow~\cite{Shimshoni1998}
\begin{equation}\begin{split}
R\propto \exp[A(\frac{H}{H_{c2}}-1)].\label{eq:3}
\end{split}\end{equation}
On the other hand, a Bose metallic (BM) phase, where the interacting Cooper pairs form a gapless non-superfluid liquid, was also proposed~\cite{Das2001,Bose-metal2}.
Based on this model, the unbinding of quantum dislocation-antidislocation
pairs due to strong gauge field fluctuations will give rise to
\begin{equation}\begin{split}
R\propto (H-H_{c0})^{2\nu_0},\label{eq:4}
\end{split}\end{equation}
where $H_{c0}$ is  the critical field for SC-BM transition. With a careful analysis, we found that Eq. (3) deviates seriously from the experimental data (not shown here). Meanwhile, Eq. (4) can describe the
experimental results in a wide field range (see the red dashed line in Fig. 3(c)). The exponent $\nu$ is determined to be 0.34. The value of critical field $H_{c0}$ (47 Oe) is quite reasonable, because the metallic state can be
induced by the lowest field in our experiment, 50 Oe. The above analysis shows that the present system is more consistent with the theoretical model based on Bose metal.

\begin{figure}
\centering
\includegraphics[width=9cm]{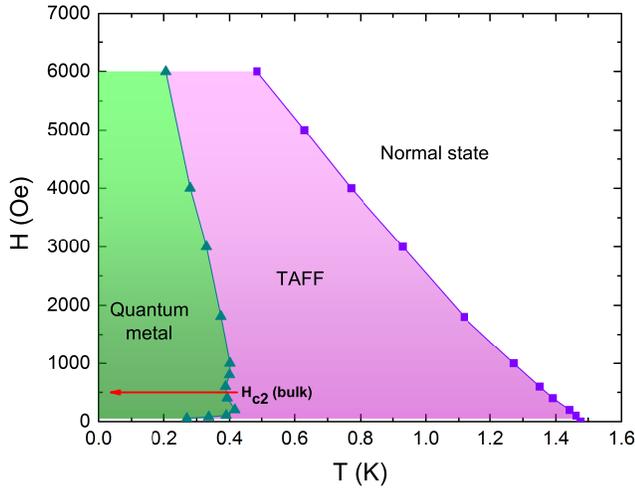}
\caption {(color online) Phase diagram for the KV$_3$Sb$_5$ thin film. The boundary between the TAFF and normal states is determined from the $R-T$ data using the criterion of 99\% $R_n$.
The boundary between the quantum metallic and TAFF states is determined by the black arrows shown in Fig. 3(a). The red arrow indicates the value of upper critical field for the bulk sample.} \label{fig4}
\end{figure}

Based on the above observations, we
can draw the field-temperature phase diagram
of the thin KV$_3$Sb$_5$, which is shown in Fig. 4. A true zero-resistance state is not observed under the lowest field of our experiment, 50 Oe. Thus we didn't indicate the SC region in this diagram. From the
magnetoresistance data in Fig. 3(c), we found that the critical field for the SC-BM transition is 47 Oe at 0.1 K. This means indirectly that there may exist a very narrow SC region under low fields.
Another remarkable feature of this
phase diagram is that the boundary between the quantum metallic and TAFF states is rather steep in a wide field range, revealing the robustness of the quantum metallic state against the magnetic field.
Finally, we would like to point out that the upper critical field exceeds that of the bulk sample (see the red arrow in Fig. 4) by more than 10 times. The origin of the such unconventional observations, especially the
possible correlation to the exotic electronic structure, is a very important topic for the theoretical realm in the future.

In summary, we successfully obtained the thin samples of the Kagome matal KV$_3$Sb$_5$ with the thickness down to 15 nm by using a mechanical exfoliation method. The dimensionality effect was observed
in terms of the higher onset SC transition temperature, 2D-featured thermal activated energy, and the BKT-type resistance transition. Moreover, the upper critical field is enhanced by more than 10 times as compared with
the bulk sample. The temperature-independent behavior of the resistance can be described by the model based on Bose metallic state.
Our results  provide a very unique and suitable platform to investigate the dimensionality effect in unconventional Kagome superconductors.


\section*{Acknowledgements}

This work is supported by the National Natural Science Foundation of China
(Nos. 11204338 and 51925208) and the Youth Innovation Promotion Association of the Chinese
Academy of Sciences (No. 2015187).


\end{document}